\documentclass[a4paper,10pt]{article}

\usepackage{amssymb,amsfonts,amsmath,epsfig,float,graphicx,caption}

\newcommand{\be}{\begin{eqnarray}}

\newcommand{\ee}{\end{eqnarray}}

\newcommand{\mailto}{\small\tt}
\def\Io{{\mathbb I}}

\def\Co{{\mathbb C}}

\def\Seff{S_{_{\rm eff}}}

\begin{document}

\title{Energy transfer using unitary transformations}

\author{
Winny O'Kelly de Galway$^1$ and Jan Naudts$^2$\\
\small{1 Instituut voor Theoretische Fysica,}\\
\small{Universiteit Leuven, K.U. Leuven, B-3001 Leuven, Belgium}\\
\small{2 Departement Fysica, Universiteit Antwerpen,}\\
\small{Universiteitsplein 1, B-2610 Antwerpen, Belgium}\\
\mailto {winny.okellydegalway@fys.kuleuven.be}, \mailto{jan.naudts@ua.ac.be}
}

\maketitle

\begin{abstract}
We study the unitary time evolution of a simple quantum Hamiltonian describing 
two harmonic oscillators coupled via a three-level system. The latter
acts as an engine transferring energy from one oscillator to the other
and is driven in a cyclic manner by time-dependent external fields.
The $S$-matrix of the cycle is obtained in analytic form.
The total number of quanta contained in the system is a conserved quantity.
As a consequence the spectrum of the $S$-matrix is purely discrete
and the evolution of the system is quasi-periodic.

\end{abstract}

% Keywords
%unitary time evolution, driven quantum system, three-level system

%%%%%%%%%%%%%%%%%%%%%%%%%%%%%%%%%%%%%%%%%%%%%%%%%%%%%%%%%%%%%%%%%%%
\section{Introduction}

The use of a three level system as an engine to transfer energy between two
quantum systems has been proposed half a century ago
by Scovil and Schulz-Dubois \cite {SSD59,GSDS67}.
The population of the levels can be manipulated using light pulses.
In particular, the Stimulated Raman Adiabatic Passage (STIRAP)
technique \cite {BTS98,VVN98,NR05}
has become a very efficient experimental tool \cite {KGPYC09}.
The three-level system is brought in contact alternatingly with 
the system of interest and with an energy reservoir, called the heat bath.
In this way energy can be removed from the system under study.

Quantum entanglement between the system and the heat bath is usually neglected.
It is assumed to be suppressed by decoherence phenomena active in the heat
bath. In the present model both the system and the reservoir consist of single harmonic
oscillators. These are too simple to cause decoherence. 
One can therefore expect that quantum entanglement is dominantly present.
The importance of the entanglement of system and reservoir
has been stressed recently \cite{ELV10}.

The thermal state of the system is usually described by a density matrix.
Here we deviate from this tradition by assuming that the state of our three component
system is described by a time-dependent wave function $\psi$ which is a solution of
the Schr\"odinger equation. It is a closed system in the sense that the
time evolution is unitary and deterministic. This corresponds experimentally
with an operation on a time scale which is short compared to the time scale
of thermal equilibration.

% overview of the paper
The model is introduced in the next Section.
The $S$-matrix approach is explained in Section 3.
The analytic expression for the $S$-matrix corresponding with one
cycle of the engine is obtained.
In Section 4 we analyze our results.
Final conclusions follow in Section 5.
The details of our calculations are explained in the Appendices.

%%%%%%%%%%%%%%%%%%%%%%%%%%%%%%%%%%%%%%%%%%%%%%%%%%%%%%%%%%%%%%%%%%%
\section{The model}

The model Hamiltonian $H$ consists of an unperturbed part $H_0$
describing two harmonic oscillators (HO) and an engine,
to which are added time-dependent external fields operating the engine and
time-dependent interactions between the oscillators and the engine.
For convenience, one of the oscillators is called the cold HO, the other the warm HO.
The engine is operated in such a way that an energy transfer from cold to warm is expected.

All together, the unperturbed Hamiltonian reads
(we use units in which $\hbar=1$)
\be
H_0=\omega_1 a^\dagger a
+ H_{gef}
+\omega_3 c^\dagger c.
\ee
The operators $a$ and $c$ are the annihilation operators of the
cold HO and of the warm HO, respectively.
The Hamiltonian of the three level system is given by
\be
H_{\rm gef}=\left(
\begin{array}{lcr}
-\mu &0 &0\\
0 &\mu &0\\
0 &0 &\mu+2\delta
\end{array}\right).
\ee
The three levels are labeled $g$, $e$, and $f$, and have energies
$-\mu$, $\mu$, and $\mu+2\delta$, respectively.

The engine is operated by means of a rather primitive sequence of two square pulses.
More realistic pulses can be treated analytically as well \cite {NOK10}
but would complicate our analysis of the coupled system as a whole.
Their contribution is
\be
I_{\rm gef}=-\epsilon_{a}(t) \Lambda_1-\epsilon_{b}(t)\Lambda_6,
\ee
where $\Lambda_1$ and $\Lambda_6$ are the Gell-Mann matrices --- see the Appendix A.

The interaction between the three level system and each of the harmonic oscillators
is inspired by the Jaynes-Cummings model. It describes an exchange of one quantum
of energy between a HO and a two-level system. Important for the present work is
that its eigenvalues and eigenvectors can be calculated analytically.

The coupling at the cold side is given by\footnote{In the Jaynes-Cummings
model $a^\dagger$ is multiplied with $\sigma_+$ instead of $\sigma_-$. The change made here
is needed because the ground state of our three level system corresponds with the excited state
in the Jaynes-Cummings model.}
\be
H_{12}=\kappa_{12}(t)\big(a^\dagger E_++ a E_-\big)
\label {model:coldvalve}
\ee
with
\be
E_+&=&\frac 12(\Lambda_1+i\Lambda_2)=\left(
\begin{array}{lcr}
0 &1 &0\\
0 &0 &0\\
0 &0 &0
\end{array}\right)
\quad\mbox{ and }\cr
E_-&=&\frac 12(\Lambda_1-i\Lambda_2)=\left(
\begin{array}{lcr}
0 &0 &0\\
1 &0 &0\\
0 &0 &0
\end{array}\right).
\ee
It couples the $g$ and $e$ levels of the three level system.
At the warm side the interaction Hamiltonian is given by
\be
H_{23}=\kappa_{23}(t)\big(F_+c^\dagger+  F_-c\big)
\label {model:warmvalve}
\ee
with
\be
F_+&=&\frac 12(\Lambda_6+i\Lambda_7)=\left(
\begin{array}{lcr}
0 &0 &0\\
0 &0 &1\\
0 &0 &0
\end{array}\right)
\quad\mbox{ and }\cr
F_-&=&\frac 12(\Lambda_6-i\Lambda_7)=\left(
\begin{array}{lcr}
0 &0 &0\\
0 &0 &0\\
0 &1 &0
\end{array}\right).
\ee
It couples the $e$ and $f$ levels of the three level system with the warm HO.
The total time-dependent Hamiltonian is now
\be
H=H_0+H_{12}+I_{\rm gef}+H_{23}.
\label{model:ham}
\ee

%%%%%%%%%%%%%%%%%%%%%%%%%%%%%%%%%%%%%%%%%%%%%%%%%%%%%%%%%%%%%%%%%%%
\section{Cycles}

\begin{figure}[ht]
 \begin{center}
 \includegraphics[width=7cm]{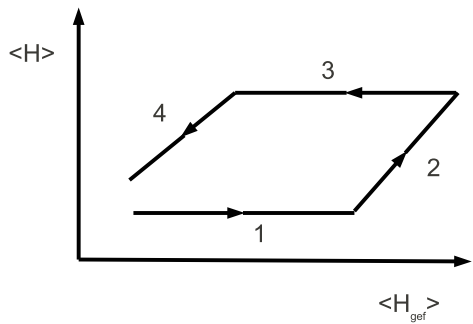}
 \end{center}
 \caption{The 4 phases of the cycle. On the horizontal axis is the energy of the engine.
On the vertical axis is the total energy of the system.}
 \label {fig:cycle}
 \end{figure}

The external field strengths $\epsilon_{a}(t)$ and $\epsilon_{b}(t)$ and the
coupling parameters $\kappa_{12}(t)$ and $\kappa_{23}(t)$ all depend on time $t$.
They are pulsed one after another in such a way that a (not necessary closed) cycle is
traversed. See the Figure \ref {fig:cycle}.

The cycle starts by coupling the engine to the cold HO.
The switching on and off changes the total
energy of the system (this contribution is omitted in the figure).
But during the first phase of the cycle the total energy is constant.
In the second phase the energy of the engine is pumped up by applying a sequence of two pulses.
Work is performed by doing so. In phase 3 the engine releases energy to the warm oscillator.
In phase 4 the engine delivers work to the environment.
This is again modeled by two externally applied pulses which
 pump down the internal energy of the engine.

\begin{figure}[ht]
 \begin{center}
  \includegraphics[width=7cm]{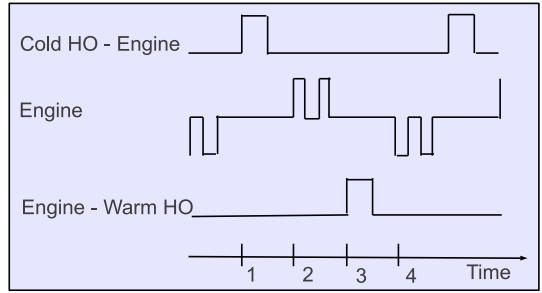}
 \end{center}
 \caption{Overview of the activation of the time-dependent terms in the Hamiltonian.}
 \label {fig:pulses}
 \end{figure}

Note that the cycle does not necessarily close.
It is obvious that in the energy transfer mode the engine will
consume more work (during phase 2) than it can deliver (during phase 4).
Because the system is finite this implies that the total energy
goes up after every cycle of the process.
%  The time inversion symmetry of the purely mechanical
% system is broken by the sequence in which external pulses are given and couplings are activated.
See the Figure \ref {fig:pulses}.

%%%%%%%%%%%%%%%%%%%%%%%%%%%%%
\subsection*{The $S$-matrix approach}

The time evolution of the system with Hamiltonian (\ref{model:ham}) is studied without making
crude approximations. The calculation is simplified by the use of the interaction picture.
Then the wave function of the total system ---  engine plus oscillators ---
is time-independent in the periods when
none of the time-dependent terms is active.
The effect of activating one of the interaction terms
or one of the external fields is then to transform the wave function $\psi$ by means of an S-matrix
into a new wave function $S\psi$.

%%%%%%%%%%%%%%%%%%%%%%%%%%%%%
\subsection*{Step 1: Absorbing energy from the cold HO}

In the first phase of the cycle the three level system is connected to the cold HO
during a time $\tau_1$. The corresponding S-matrix is denoted $S_1$.
It is not very difficult to calculate it exactly. See the Appendix B.
The result is of the form
\be
S_1
&=&e^{i\tau_{1}H_0}e^{-i\tau_{1}H}\cr
&=&e^{\frac i2\tau_1(\omega_1-2\mu)}\left[
a^\dagger (A-iC)a E_1+ia^\dagger B E_+\right]\cr
& &+e^{-\frac i2\tau_1(\omega_1-2\mu)}\left[
iBa E_- +aa^\dagger (A+iC) E_2\right]\cr
& & +G_1E_1+E_3,
\label {S:step1}
\ee
with
\be
E_1&=&E_+E_-=\left(\begin{array}{lcr}
1 &0 &0\\
0 &0 &0\\
0 &0 &0\\
\end{array}\right),
\quad
E_2=E_-E_+=\left(\begin{array}{lcr}
0 &0 &0\\
0 &1 &0\\
0 &0 &0\\
\end{array}\right),
\cr
E_3&=&\left(\begin{array}{lcr}
0 &0 &0\\
0 &0 &0\\
0 &0 &1\\
\end{array}\right),
\ee
and
\be
A&=&\sum_n\frac 1{n+1}\cos(\tau_1\lambda_n)|n\rangle\langle n|,\cr
B&=&\sum_n\frac 1{\sqrt{n+1}}\sin(\tau_1\lambda_{n})\sin(2\theta_{n})|n\rangle\langle n|,\cr
C&=&\sum_n\frac 1{n+1}\sin(\tau_1\lambda_n)\cos(2\theta_n)|n\rangle\langle n|.\cr
& &
\ee
The coefficients $\lambda_n$ and the angles $\theta_n$ are given by
\be
\lambda_n&=&\frac 12\sqrt{4\kappa_{12}^2(n+1)+(\omega_1-2\mu)^2}\cr
\tan(\theta_n)&=&\frac {2\kappa_{12}\sqrt{n+1}}{2\lambda_n+\omega_1-2\mu}.
\ee
The operator $G_1$ is the orthogonal projection $|0\rangle\langle 0|$
onto the ground state of the cold HO.

%%%%%%%%%%%%%%%%%%%%%%%%%%%%%
\subsection*{Step 2: Pumping up}

We apply a sequence of two pulses of the on/off type.
The first pulse is realized by giving $\epsilon_{a}(t)$ a constant non-zero value
during a time $\tau_{a}$. It tries to invert the population of the levels $e$ and $f$.
The change of the population as a consequence of
this pulse is given by the S-matrix $S_{2a}$ which is now calculated.
\be
S_{2a}&=&e^{i\tau_{a}H_0}e^{-i\tau_{a}H}\cr
&=&e^{i\tau_{a}H_{\rm gef}}e^{-i\tau_{a}(H_{\rm gef}-\epsilon_{a}\lambda_6)}\cr
&=&e^{- i\tau_{a}\delta\sigma_3}
e^{ i\tau_{a}[\delta\sigma_3-\epsilon_{a}\sigma_1]}.
\ee
Note that we switched notations,
 using two-dimensional Pauli matrices instead of the Gell-Mann matrices,
omitting one dimension for a moment.
Introduce the constant $T_{a}=1/\sqrt{\delta^2+\epsilon_{a}^2}$. There follows
\be
S_{2a}&=&
\left[\cos(\tau_{a}\delta)-i\sin(\tau_{a}\delta)\sigma_3\right]\cr
& &\times
\left[\cos(\tau_{a}/T_{a})+iT_{a}\sin (\tau_{a}/T_{a})(\delta\sigma_3-\epsilon_{a}\sigma_1)\right].\cr
& &
\ee

Let us now make an appropriate choice of the pulse duration $\tau_{a}$.
The goal is to minimize the population of the $e$-level after the pulse.
Since one can expect that before the pulse the $e$-level is more populated than the $f$-level
the best one can do is to require that the $e$ matrix element
 of $S_{2a}$ is as small as possible in modulus.
Let therefore $\tau_{a}=\frac 12\pi T_{a}$. Then the S-matrix becomes
\be
S_{2a}&=&iT_{a}\left[\cos(\tau_{a}\delta)-i\sin(\tau_{a}\delta)\sigma_3\right]
\sin (\tau_{a}/T_{a})(\delta\sigma_3-\epsilon_{a}\sigma_1).
\ee
Restoring the third dimension this becomes
\be
S_{2a}
&=&\left(\begin{array}{lcr}
		  1 &0 &0\\
		  0 &0 &0\\
		  0 &0 &0
                \end{array}\right)
+iT_{a}
\left(\begin{array}{lcr}
		  0 &0 &0\\
		  0 &\delta e^{-i\tau_{a}\delta} &-\epsilon_{a}e^{-i\tau_{a}\delta}\\
		  0 &-\epsilon_{a}e^{i\tau_{a}\delta} &-\delta e^{i\tau_{a}\delta}
                \end{array}\right).\cr
& &
\ee
In the limit of a strong short pulse this becomes
\be
S_{2a}
&=&
\left(\begin{array}{lcr}
		  1 &0 &0\\
		  0 &0 &-i\\
		  0 &-i &0
                \end{array}\right).
\label{cycles:step2simpl}
\ee

The first pulse of the second phase of the cycle is followed by a pulse of duration
$\tau_{b}$, intended to invert the population of levels $e$ and $g$.
The corresponding S-matrix reads,
using the notation $T_{b}=1/\sqrt{\mu^2+\epsilon_{b}^2}$,
\be
S_{2b}
&=&e^{i\tau_{b}H_0}e^{-i\tau_{b}H}\cr
&=&e^{i\tau_{b}H_{\rm gef}}e^{-i\tau_{b}(H_{\rm gef}-\epsilon_{b} \lambda_1)}\cr
&=&\left[\cos(\tau_{b}\mu)-i\sin(\tau_{b}\mu)\sigma_3\right]\cr
& &\times
\left[\cos(\tau_{b}/T_{b})+iT_{b}\sin (\tau_{b}/T_{b})(\mu\sigma_3+\epsilon_{b}\sigma_1\right]\cr
& &
\ee
With similar arguments as before let us choose $\tau_{b}=\frac 12\pi T_{b}$.
Then the S-matrix becomes
\be
S_{2b}
&=&\left(\begin{array}{lcr}
		  0 &0 &0\\
		  0 &0 &0\\
		  0 &0 &1
                \end{array}\right)
+iT_{b}\left(\begin{array}{lcr}
		  \mu e^{-i\tau_{b}\mu} &\epsilon_{b}e^{-i\tau_{b}\mu}  &0\\
		  \epsilon_{b}e^{i\tau_{b}\mu} &-\mu e^{i\tau_{b}\mu} &0\\
		  0 &0 &0
                \end{array}\right).
\ee
In the limit of a strong short pulse this becomes
\be
S_{2b}
&=&
\left(\begin{array}{lcr}
		  0 &i  &0\\
		  i &0 &0\\
		  0 &0 &1
                \end{array}\right).
\ee

All together the S-matrix for the second phase of the cycle equals
\be
S_2&=&S_{2b}S_{2a}\cr
&=&\left(\begin{array}{lcr}
  iT_b\mu e^{-i\tau_b\mu} &-T_aT_b\epsilon_b\delta e^{-i\tau_a\delta-i\tau_{b}\mu}  &T_aT_b\epsilon_a\epsilon_be^{-i\tau_a\delta-i\tau_{b}\mu}\\
  iT_b\epsilon_be^{i\tau_{b}\mu} &T_aT_b\mu\delta e^{-i\tau_a\delta+i\tau_{b}\mu} &-T_aT_b\mu\epsilon_ae^{-i\tau_a\delta+i\tau_{b}\mu}\\
  0 &-iT_a\epsilon_a e^{i\tau_a\delta}&-iT_a\delta e^{-i\tau_a\delta}
         \end{array}\right).\cr
& &
\ee
In the limit of strong short pulses it becomes
\be
S_{2}
&=&
\left(\begin{array}{lcr}
		  0 &0 &1\\
		  i &0 &0\\
		  0 &-i &0
                \end{array}\right).
\label {S2limit}
\ee

%%%%%%%%%%%%%%%%%%%%%%%%%%%%%
\subsection*{Step 3: Exchanging energy with the warm oscillator}

In the third phase of the cycle the three level system is connected to the warm HO
during a time $\tau_3$. The corresponding S-matrix is denoted $S_3$.
The calculation is similar to that in Step 1. The result is of the form
\be
S_3
&=&e^{i\tau_{3}H_0}e^{-i\tau_{3}H}\cr
&=&E_1+E_2G_3\cr
& &+e^{\frac i2\tau_3(\omega_3-2\delta)}\left[
c^\dagger (Z-iV)c E_2+ic^\dagger Y F_+\right]\cr
& &+e^{-\frac i2\tau_3(\omega_3-2\delta)}\left[
iYc F_- +cc^\dagger (Z+iV) E_3\right]
\ee
with
\be
Z&=&\sum_n\frac 1{n+1}\cos(\tau_3\xi_n)|n\rangle\langle n|\cr
Y&=&\sum_n\frac 1{\sqrt{n+1}}\sin(\tau_3\xi_{n})\sin(2\phi_{n})|n\rangle\langle n|,\cr
V&=&\sum_n\frac 1{n+1}\sin(\tau_3\xi_n)\cos(2\phi_n)|n\rangle\langle n|.\cr
& &
\ee
The coefficients $\xi_n$ and the angles $\phi_n$ are given by
\be
\xi_n&=&\frac 12\sqrt{4\kappa_{23}^2(n+1)+(\omega_3-2\delta)^2}\cr
\tan(\phi_n)&=&\frac {2\kappa_{23}\sqrt{n+1}}{2\xi_n+\omega_3-2\delta}.
\ee
The operator $G_3$ is the orthogonal projection $|0\rangle\langle 0|$
 onto the ground state of the warm HO.

%%%%%%%%%%%%%%%%%%%%%%%%%%%%%
\subsection*{Step 4: Pumping down}

The operation in the fourth phase is
the inverse of that in the second phase. We thus have $S_4=S_2^\dagger$.

%%%%%%%%%%%%%%%%%%%%%%%%%%%%%%%%%%%%%%%%%%%%%%%%%%%%%%%%%%%%%%%%%%%
\section{Analysis}

In the previous Section the contribution to the S-matrix from each
 of the four phases of the cycle has been obtained.
The composite matrix $S=S_4S_3S_2S_1$ is now calculated. The result is a rather complicated.
Therefore a tensor notation is appropriate. Remem\-ber that the Hilbert space of wave functions of
the total system is the tensor product
\be
{\cal H}={\cal H}_{\rm cold}\otimes\Co^3\otimes{\cal H}_{\rm warm}.
\ee
The first and the last factor are the Hilbert space of the cold and of the warm
HO, respectively.
The middle factor is the space of vectors with three complex components.

%%%%%%%%%%%%%%%%%%%%%%%%%%%%%
\subsection{The composed S-matrix}

The full S-matrix reads
\be
S
&=&\Io\otimes S_2^\dagger  \otimes\Io\cr
&\times&  \bigg\{
\Io\otimes E_1\otimes\Io+\Io\otimes E_2\otimes G_3\cr
& &+e^{\frac i2\tau_3(\omega_3-2\delta)}\left[
\Io\otimes E_2\otimes c^\dagger (Z-iV)c+i\Io\otimes F_+\otimes c^\dagger Y \right]\cr
& &+e^{-\frac i2\tau_3(\omega_3-2\delta)}\left[
i\Io\otimes F_-\otimes Yc +\Io\otimes E_3\otimes cc^\dagger (Z+iV)\right]
\bigg\}\cr
&\times& \Io\otimes S_2\otimes\Io\cr
&\times&  \bigg\{e^{\frac i2\tau_1(\omega_1-2\mu)}\left[
a^\dagger (A-iC)a\otimes E_1\otimes\Io+ia^\dagger B\otimes E_+\otimes\Io\right]\cr
& &+e^{-\frac i2\tau_1(\omega_1-2\mu)}\left[
iBa \otimes E_-\otimes\Io +aa^\dagger (A+iC)\otimes E_2\otimes\Io\right]\cr
& &+G_1\otimes E_1\otimes\Io +\Io\otimes E_3\otimes\Io\bigg\}\cr
&=& \bigg\{
\Io\otimes S_2^\dagger E_1 S_2\otimes\Io+\Io\otimes S_2^\dagger E_2S_2\otimes G_3\cr
& &+e^{\frac i2\tau_3(\omega_3-2\delta)}\left[
\Io\otimes S_2^\dagger E_2 S_2\otimes c^\dagger (Z-iV)c+i\Io\otimes S_2^\dagger F_+ S_2\otimes c^\dagger Y \right]\cr
& &+e^{-\frac i2\tau_3(\omega_3-2\delta)}\left[
i\Io\otimes S_2^\dagger F_- S_2\otimes Yc +\Io\otimes S_2^\dagger E_3 S_2\otimes cc^\dagger (Z+iV)\right]
\bigg\}\cr
&\times&  \bigg\{e^{\frac i2\tau_1(\omega_1-2\mu)}\left[
a^\dagger (A-iC)a\otimes E_1\otimes\Io+ia^\dagger B\otimes E_+\otimes\Io\right]\cr
& &+e^{-\frac i2\tau_1(\omega_1-2\mu)}\left[
iBa \otimes E_-\otimes\Io +aa^\dagger (A+iC)\otimes E_2\otimes\Io\right]\cr
& &+G_1\otimes E_1\otimes\Io +\Io\otimes E_3\otimes\Io\bigg\}
\ee
For simplicity, we use the value  (\ref {S2limit}) of $S_2$ in the limit of strong short pulses.
In this limit one has
$S_2^\dagger E_1S_2=E_3$,
$S_2^\dagger E_2S_2=E_1$,
$S_2^\dagger E_3S_2=E_2$,
$S_2^\dagger F_+S_2=-E_+$,
$S_2^\dagger F_-S_2=-E_-$.
Hence, the above expression for $S$ simplifies to
\be
S&=& \bigg\{
\Io\otimes E_3\otimes\Io+\Io\otimes E_1\otimes G_3\cr
& &+e^{\frac i2\tau_3(\omega_3-2\delta)}\left[
\Io\otimes E_1\otimes c^\dagger (Z-iV)c-i\Io\otimes E_+\otimes c^\dagger Y \right]\cr
& &+e^{-\frac i2\tau_3(\omega_3-2\delta)}\left[
-i\Io\otimes E_-\otimes Yc +\Io\otimes E_2 \otimes cc^\dagger (Z+iV)\right]
\bigg\}\cr
&\times&  \bigg\{e^{\frac i2\tau_1(\omega_1-2\mu)}\left[
a^\dagger (A-iC)a\otimes E_1\otimes\Io+ia^\dagger B\otimes E_+\otimes\Io\right]\cr
& &+e^{-\frac i2\tau_1(\omega_1-2\mu)}\left[
iBa \otimes E_-\otimes\Io +aa^\dagger (A+iC)\otimes E_2\otimes\Io\right]\cr
& & +G_1\otimes E_1\otimes\Io+\Io\otimes E_3\otimes\Io\bigg\}\cr
&=&\Io\otimes E_3\otimes\Io+G_1\otimes E_1\otimes G_3\cr
& &+e^{\frac i2\tau_1(\omega_1-2\mu)}\left[
a^\dagger (A-iC)a\otimes E_1\otimes G_3+ia^\dagger B\otimes E_+\otimes G_3\right]\cr
& &+e^{\frac i2\tau_3(\omega_3-2\delta)}
G_1\otimes E_1\otimes c^\dagger (Z-iV)c
-ie^{-\frac i2\tau_3(\omega_3-2\delta)}
G_1\otimes E_-\otimes Yc\cr
& &+e^{\frac i2\tau_3(\omega_3-2\delta)}e^{\frac i2\tau_1(\omega_1-2\mu)}\cr
& &\quad\times
\left[a^\dagger (A-iC)a\otimes E_1
+ia^\dagger B\otimes E_+\right]\otimes c^\dagger (Z-iV)c\cr
& &+e^{\frac i2\tau_3(\omega_3-2\delta)}e^{-\frac i2\tau_1(\omega_1-2\mu)}
\left[Ba\otimes E_1
-iaa^\dagger (A+iC)\otimes E_+\right]\otimes c^\dagger Y\cr
& &+e^{-\frac i2\tau_3(\omega_3-2\delta)}e^{\frac i2\tau_1(\omega_1-2\mu)}
\left[-ia^\dagger (A-iC)a\otimes E_-
+a^\dagger B\otimes E_2\right]\otimes Yc\cr
& &+e^{-\frac i2\tau_3(\omega_3-2\delta)}e^{-\frac i2\tau_1(\omega_1-2\mu)}\cr
& &\quad\times
\left[iBa\otimes E_-
+aa^\dagger (A+iC)\otimes E_2\right]\otimes cc^\dagger (Z+iV).\cr
& &
\label {anal:Sres}
\ee
Note that the operators $A$, $B$, $C$, $Y$, $Z$, $V$, commute with the counting operators of the two
harmonic oscillators. Hence the two terms which directly transfer energy between the two
oscillators are those proportional to $Ba\otimes E_1\otimes c^\dagger Y$ and
$a^\dagger B\otimes E_2\otimes Yc$ respectively. They act in opposite directions.
Other terms do not transfer energy or they exchange energy between
 the engine and one of the oscillators.
See the Table \ref {anal:table}.

\begin{table}[t!]
\caption{\label{anal:table}
Interpretation of the terms appearing in (\ref {anal:Sres}).\\
The arrows indicate the direction of the energy flow, between the cold HO and the engine,
between the engine and the warm HO, respectively.}
%\begin{indented}
\begin{itemize}
\item[]\begin{tabular}{@{}lcc}
%\br
\hline
 & &\\
%\mr
$a^\dagger (A-iC)a\otimes E_1\otimes c^\dagger (Z-iV)c$		&--- 		&---\\
$a^\dagger B\otimes E_+\otimes c^\dagger (Z-iV)c$		&$\leftarrow$	&---\\
$Ba\otimes E_1\otimes c^\dagger Y$				&$\rightarrow$	&$\rightarrow$\\
$aa^\dagger (A+iC)\otimes E_+\otimes c^\dagger Y$		&---		&$\rightarrow$\\
$a^\dagger (A-iC)a\otimes E_-\otimes Yc$			&---		&$\leftarrow$\\
$a^\dagger B\otimes E_2\otimes Yc$				&$\leftarrow$	&$\leftarrow$\\
$Ba\otimes E_-\otimes cc^\dagger (Z+iV)$			&$\rightarrow$	&---\\
$aa^\dagger (A+iC)\otimes E_2\otimes cc^\dagger (Z+iV)$		&---		&---\\
%\br
\hline
\end{tabular}
%\end{indented}
\end{itemize}

\end{table}

%%%%%%%%%%%%%%%%%%%%%%%%%%%%%
\subsection{Eigenvectors of the S-matrix}

The above S-matrix describes the effect in the interaction picture of performing one cycle.
It is immediately clear that the ground state $|0,g,0\rangle$ of the system is an
eigenstate of this S-matrix with eigenvalue 1. This is an immediate consequence of the fact that
the ground state of the Jaynes-Cummings model is not affected by the interactions of the model.
An important question is whether the S-matrix has other eigenvectors. Indeed, such eigenvectors
describe situations in which the action of the engine has no effect at all.
Of course, on a superposition of eigenvectors the engine can have effect.
But the result is an almost periodic function which always returns arbitrary close to its
starting point.
On the other hand, if part of the spectrum of $S$ is continuous then a genuine energy transfer is
possible by which the system approaches a stationary regime.

An easy argument shows that the spectrum of the S-matrix is purely discrete.
The Jaynes-Cummings interaction term describes the exchange of a single quantum of energy
between a HO and a two-level system. The external action onto the three-level engine
changes the total energy of the system but not the number of quanta it contains.
As a consequence the Hilbert space of wave functions $\cal H$ decomposes into finite dimensional
subspaces ${\cal H}_n$ containing an exact number $n$ of quanta.
Indeed, the subspace ${\cal H}_n$ is generated by the $2n+1$ basis vectors
\be
& &|m,g,n-m\rangle, m=0,\cdots,n,\cr
\mbox{ and }
& &
|m,e,n-m-1\rangle,m=0,\cdots,n-1.
\ee
Using the explicit expression (\ref {anal:Sres}) one verifies that ${\cal H}_n$
is invariant under $S$.

%%%%%%%%%%%%%%%%%%%%%%%%%%%%%
\subsection{Energy transfer}

The result (\ref {anal:Sres}) seems hopelessly complicated but can never the less
be used to derive some unexpected properties of the engine.
The change in the energy of the cold HO before and after one cycle 
is defined by
\be
D=S^\dagger a^\dagger aS-a^\dagger a.
\ee
One finds (see the Appendix C)
\be
D
&=&\bigg[aa^\dagger B^2\otimes E_2-a^\dagger B^2a\otimes E_1\cr
& &+ia^\dagger  aa^\dagger (A+iC)B\otimes E_+
-i(A-iC)B aa^\dagger a\otimes E_-\bigg]\otimes\Io.
\label {transfer:en1}
\ee
The eigenvectors of $D$ are of the form
\be
\psi=u|n+1\rangle\otimes|g\rangle+v|n\rangle\otimes |e\rangle
\ee
(we neglect the Hilbert space of the warm HO for a moment).
The condition $D\psi=\rho\psi$ then yields
\be
0&=&u\left(\rho+\sin^2(\tau_1\lambda_n)\sin^2(2\theta_n)\right)\cr
& &+v\sin(\tau_1\lambda_n)\sin(2\theta_n)\left[\sin(\tau_1\lambda_n)\cos(2\theta_n)-i\cos(\tau_1\lambda_n)\right]\cr
0&=&u\sin(\tau_1\lambda_n)\sin(2\theta_n)\left[\sin(\tau_1\lambda_n)\cos(2\theta_n)+i\cos(\tau_1\lambda_n)\right]\cr
& &+v\left(\rho-\sin^2(\tau_1\lambda_n)\sin^2(2\theta_n)\right).
\ee
This set of equations has a non-trivial solution when
\be
\rho=\pm \sin(\tau_1\lambda_n)\sin(2\theta_n).
\ee
Corresponding eigenvectors are then
\be
u&=&\sin(\tau_1\lambda_n)\cos(2\theta_n)-i\cos(\tau_1\lambda_n),\cr
v&=&\mp 1-\sin(\tau_1\lambda_n)\sin(2\theta_n).
\ee

Note that $D|0\rangle\otimes |g\rangle=0$. Hence, the spectrum of $D$ is completely known.
For each strictly positive eigenvalue $\rho>0$ also $-\rho$ is an eigenvalue.
$\rho>0$ corresponds with raising the energy of the cold HO, $\rho<0$ with cooling.

One concludes that raising or lowering the energy of the cold HO
after one cycle of the engine depends completely on the choice of the initial wave function.
The important question is of course what happens after one cycle with a wave function
originally chosen as an eigenvector $\psi$ of $D$ with negative eigenvalue. Will $S\psi$
be a superposition of eigenvectors all with negative eigenvalues? Or will part of them
have a positive eigenvalue? 
Preliminary numerical evaluations show that the latter is the case.
The resulting behavior of the engine is rather complicated.

A similar calculation for the warm HO is possible.
But note that an easy result only follows when starting the cycle with
coupling the engine to the warm HO
instead of the cold HO, as used in the above calculations.

%%%%%%%%%%%%%%%%%%%%%%%%%%%%%
\subsection{Performing work}

The previous subsections give a partial answer to the question whether the engine is capable
of transferring energy between the two oscillators. Now follows a discussion of the work
needed to operate the engine.

In phases 1 and 3 of the cycle some work is needed to operate the
valves connecting the engine with the cold HO respectively the warm HO.
Indeed, switching on and off the interaction terms (\ref {model:coldvalve}, \ref {model:warmvalve})
changes the total energy of the system. Since the wave function of the system evolves in time between
the switching on and switching off the involved energy changes to not necessarily cancel.
Hence we expect that a tiny amount of work is needed to operate these valves.
% We do not calculate this contribution but concentrate on the major contribution coming from phases
% 2 and 4 of the cycle.

It is now indicated to consider a cycle starting with phase 2 instead of phase 1.
Then the energy changes during the respective phases are given by
\be
\Delta E_1&=&H_0-S_1H_0S_1^\dagger\cr
\Delta E_2&=&S_2^\dagger H_0S_2-H_0,\cr
\Delta E_3&=&S_2^\dagger(S_3^\dagger H_0S_3-H_0)S_2,\cr
\Delta E_4&=&S_2^\dagger S_3^\dagger(S_2 H_0S_2^\dagger-H_0)S_3S_2.
\ee
Using the simplified expression (\ref {cycles:step2simpl}) for $S_2$ one obtains
\be
\Delta E_1&=&
(\omega_1-2\mu)\left(B^2aa^\dagger E_2-a^\dagger B^2a E_1\right)\cr
& &+i(\omega_1-2\mu)a^\dagger Baa^\dagger (A+iC)E_+\cr
& &-i(\omega_1-2\mu) Baa^\dagger(A-iC)a E_-
\label {work:phase1}
\ee
and
\be
\Delta E_2&=&2[\mu,\delta,-\mu-\delta]
\ee
and
\be
\Delta E_3&=&(\omega_3-2\delta)
\left[E_2cc^\dagger Y^2 -E_1c^\dagger Y^2 c\right]\cr
& &-i(\omega_3-2\delta)
E_+c^\dagger (Z+iV)cc^\dagger Y\cr
& &+i(\omega_3-2\delta)E_-(Z-iV)cc^\dagger Yc
\label {work:phase3}
\ee
and
\be
\Delta E_4
&=&-\Delta E_2+2(\mu-\delta)\left(E_1 c^\dagger Y^2 c-E_2 Y^2cc^\dagger\right)\cr
& &+2i(\mu-\delta)\left[E_+ c^\dagger(Z+iV)cc^\dagger Y
-E_- Ycc^\dagger(Z-iV)c\right],\cr
& &
\label {work:phase4}
\ee
where $[a,b,c]$ denotes the diagonal matrix with eigenvalues $a,b,c$.
See the Appendix D.

Several features can be observed. The contributions $\Delta E_3$ and
$\Delta E_1$ represent the energy needed to switch on and off the
interactions with the harmonic oscillators. They vanish when the coupling
between the engine and the oscillators is at resonance.

The work performed by the engine equals the quantum expectation of the operator
$-\Delta E_2-\Delta E_4$.
When $\mu=\delta$ then the operation of the engine is
meaningless and no net energy is used and no net work is performed during
the phases 2 and 4. In the general case the eigen values of $\Delta E_2+\Delta E_4$
can be calculated analytically. One obtains
\be
\lambda=\pm \sin(\tau_3\xi_n)\sin(2\phi_n).
\ee
The corresponding eigen vectors are linear combinations of $|g,n+1\rangle$
and $|e,n\rangle$ (neglecting the state of the cold HO).
Hence also the spectrum of this operator is symmetric under
a change of sign. This means that the initial conditions determine whether
operating the engine consumes energy or whether it performs work.

%%%%%%%%%%%%%%%%%%%%%%%%%%%%%
\subsection{Effective S-matrix}

Introduce the unitary operator
\be
\Seff=\left(\begin{array}{lr}
            G_1+a^\dagger(A-iC)a &ia^\dagger B\\
	    iBa         &aa^\dagger(A+iC)
            \end{array}\right)\otimes\Io.
\ee
To verify that $\Seff^\dagger\Seff=\Seff\Seff^\dagger=\Io$ use that
\be
aa^\dagger B^2+(aa^\dagger)^2(A^2+C^2)=\Io
\label{eff:id1}
\ee
and
\be
G_1+a^\dagger B^2a+a^\dagger a a^\dagger(A^2+C^2)a=\Io.
\ee
One calculates
\be
\Seff^\dagger a^\dagger a\Seff
&=&\left(\begin{array}{lr}
            a^\dagger(1-B^2)a &ia^\dagger (A+iC)aa^\dagger B\\
	    iBaa^\dagger (A-iC)a         &a^\dagger a +aa^\dagger B^2
            \end{array}\right)\otimes\Io\cr
&=&D+a^\dagger a\cr
&=&S^\dagger a^\dagger aS.
\ee
This shows that in the definition of $D$ one can use $\Seff$ instead of $S$.

%%%%%%%%%%%%%%%%%%%%%%%%%%%%%%%%%%%%%%%%%%%%%%%%%%%%%%%%%%%%%%%%%%%
\section{Conclusions}

It is feasible to obtain analytic results for a closed quantum system
consisting of an engine operating between two small quantum systems,
{\sl in casu} two harmonic oscillators.
The engine is operated by switching external fields on and off.
The state of the system is at any moment determined by
its wave function. The time evolution follows by solving the
Schr\"odinger equation using a time-dependent Hamiltonian.

In the traditional approach one considers a heat engine operating between
the system of interest and a heat bath.
The heat bath belongs to the environment and is taken
into account in a phenomenological way. The present paper considers a closed
system. Its state is described by a time-dependent wave function.
The time evolution is unitary and the quantum entanglement between the engine and the
two harmonic oscillators is treated rigorously.

From our toy model we have learned a number of points.
\begin{itemize}
\item The use of the interaction picture
improves the transparency of the calculations.
\item We do not make use of the adiabatic theorem. The change in the
population of the energy levels of the engine results from the time
evolution. As a consequence all results depend only on intra-level
distances and not on the positioning of oscillator levels 
w.r.t.~levels of the engine.
\item At each of the two interfaces the energy flows in both directions.
Energy leaks away in the direction opposite to the intended one.
Eight different energy contributions have been distinguished in
the Table \ref {anal:table}. In the usual approach these are replaced
by two phenomenological terms.

\item The $S$-matrix of a single cycle of the engine has a
purely discrete spectrum. This follows immediately from the observation that
the number of energy quanta in the system is conserved.
The total energy is not conserved. The engine changes the energy content of a quantum
before passing it on to one of the harmonic oscillators.
\item The operator $D=S^\dagger a^\dagger aS-a^\dagger a$
which measures the change in energy of the cold harmonic oscillator during one cycle
of the engine 
has a fully discrete spectrum with explicitly known eigenvectors and eigenvalues.
This is a benefit of using the Jaynes-Cummings mechanism for the interactions
between the engine and the harmonic oscillators.
\item
The spectrum of this operator $D$ is symmetric under the change of sign.
This could be a more general feature being a consequence of time inversion symmetry.
\item The change of energy of the system as a whole during one cycle
can be obtained analytically as well.
The operation of the valves connecting the engine with the oscillators costs
energy except when the interaction is at resonance. The pumping up and down of the
occupational probabilities of the engine levels can cost energy or can perform work
depending on the initial state of the system, this is, depending on its wave function.
This shows that the engine can be used either to transfer energy from the cold to
the warm oscillator or to perform work produced by the energy flow from warm to cold.
\end{itemize}

Knowing the S-matrix for a single cycle in an analytic form
makes it possible to do easy and accurate numerical simulations of many consecutive cycles.
Preliminary results show that the energy transfer is feasible.
They also show that an initial product state gets rapidly entangled
to a high and fairly constant level.
A full report of the numerical work will be published elsewhere.

\appendix
\section{The Gell-Mann matrices}

Conventionally, the Gell-Mann matrices are defined as follows.
\be
\Lambda_1&=\left(\begin{array}{lcr}
0 &1 &0\\
1 &0 &0\\
0 &0 &0\\
\end{array}\right),
\quad
\Lambda_2&=\left(\begin{array}{lcr}
0 &-i &0\\
i &0 &0\\
0 &0 &0\\
\end{array}\right),\crcr
\Lambda_3&=\left(\begin{array}{lcr}
1 &0 &0\\
0 &-1 &0\\
0 &0 &0\\
\end{array}\right),
\quad
\Lambda_4&=\left(\begin{array}{lcr}
0 &0 &1\\
0 &0 &0\\
1 &0 &0\\
\end{array}\right),\crcr
\Lambda_5&=\left(\begin{array}{lcr}
0 &0 &-i\\
0 &0 &0\\
i &0 &0\\
\end{array}\right),
\quad
\Lambda_6&=\left(\begin{array}{lcr}
0 &0 &0\\
0 &0 &1\\
0 &1 &0\\
\end{array}\right),\crcr
\Lambda_7&=\left(\begin{array}{lcr}
0 &0 &0\\
0 &0 &-i\\
0 &i &0\\
\end{array}\right),
\quad
\Lambda_8&=\frac 1{\sqrt 3}\left(\begin{array}{lcr}
1 &0 &0\\
0 &1 &0\\
0 &0 &-2\\
\end{array}\right).
\ee

\section{The S-matrix of phase 1 of the cycle}

Here we calculate the S-matrix of a Jaynes-Cummings Hamiltonian in which the interaction
is switched on during a finite time.
The coupling is constant with strength $\kappa_{12}$ during a time interval of length $\tau_1$.
The relevant Hamiltonian is
\be 
H=\omega_1 a^\dagger a-\mu\sigma_z-\kappa_{12}(a^\dagger\sigma_++a\sigma_-).
\ee
Let
\be
|g\rangle\equiv \left(\begin{array}{c} 1\\0 \end{array}\right)
\quad\mbox{ and }\quad
|e\rangle\equiv \left(\begin{array}{c} 0\\1 \end{array}\right).
\ee
The eigenstates of the HO are denoted $n\rangle$, with $n=0,1,\cdots$.

The ground state of $H$ is
\be
|0,-\rangle\equiv |0\rangle\otimes|g\rangle.
\ee
It satisfies $H|0,-\rangle=-\mu|0,-\rangle$.
The pairs of excited states are denoted $|n,\pm\rangle$, with $n=0,1,\cdots$.
They are of the form
\be
|n,-\rangle&=&\cos(\theta_n) |n\rangle\otimes|e\rangle+\sin(\theta_n) |n+1\rangle\otimes |g\rangle,\cr
|n,+\rangle&=&-\sin(\theta_n) |n\rangle\otimes|e\rangle+\cos(\theta_n) |n+1\rangle\otimes |g\rangle.
\ee
From
\be
H|n\rangle\otimes |e\rangle&=&(n\omega_1+\mu)|n\rangle\otimes |e\rangle-\kappa_{12}\sqrt{n+1}|n+1\rangle\otimes |g\rangle\cr
H|n\rangle\otimes |g\rangle&=&(n\omega_1-\mu)|n\rangle\otimes |g\rangle-\kappa_{12}\sqrt{n}|n-1\rangle\otimes |e\rangle
\ee
follows
\be
H|n,-\rangle&=&\cos(\theta_n)\left[(n\omega_1+\mu)|n\rangle\otimes|e\rangle-\kappa_{12}\sqrt{n+1}|n+1\rangle\otimes |g\rangle\right]\cr
&+&\sin(\theta_n)\left[((n+1)\omega_1-\mu) |n+1\rangle\otimes |g\rangle-\kappa_{12}\sqrt{n+1}|n\rangle\otimes |e\rangle\right],\cr
H|n,+\rangle&=&-\sin(\theta_n)\left[ (n\omega_1+\mu)|n\rangle\otimes |e\rangle-\kappa_{12}\sqrt{n+1}|n+1\rangle\otimes |g\rangle\right]\cr
&+&\cos(\theta_n)\left[((n+1)\omega_1-\mu)|n+1\rangle\otimes |g\rangle-\kappa_{12}\sqrt{n+1}|n\rangle\otimes |e\rangle\right].\cr
& &
\ee
The requirement that $H|n,\pm\rangle=E_n^\pm|n,\pm\rangle$ then yields the set of equations
\be
(n\omega_1+\mu)\cos(\theta_n)   -\kappa_{12}\sqrt{n+1}\sin(\theta_n)&=&E_n^-\cos(\theta_n)\cr
-\kappa_{12}\sqrt{n+1}\cos(\theta_n) +((n+1)\omega_1-\mu)\sin(\theta_n)&=&E_n^-\sin(\theta_n)\cr
-\kappa_{12}\sqrt{n+1}\cos(\theta_n) -(n\omega_1+\mu)\sin(\theta_n)&=&-E_n^+\sin(\theta_n)\cr
((n+1)\omega_1-\mu)\cos(\theta_n)+\kappa_{12}\sqrt{n+1}\sin(\theta_n)&=&E_n^+\cos(\theta_n)\cr
& &
\ee
The solution is
\be
E_n^\pm&=&\left(n+\frac 12\right)\omega_1\pm\lambda_n\cr
\tan\theta_n&=&\frac {2\lambda_n+2\mu-\omega_1}{2\kappa_{12}\sqrt{n+1}}
\ee
with
\be
\lambda_n=\sqrt{\kappa_{12}^2(n+1)+\left(\mu-\frac 12\omega_1\right)^2}.
\ee
A short calculation now gives
\be
S_1
|n\rangle\otimes |e\rangle&=&e^{itH_0}e^{-itH}|n\rangle\otimes |e\rangle\cr
&=&e^{itH_0}\left[\cos(\theta_n)e^{-itE_n^-}|n,-\rangle
-\sin(\theta_n)e^{-itE_n^+}|n,+\rangle\right]\cr
&=&e^{it(\mu-\omega_1/2)}\left[\cos(t\lambda_n)+i\cos(2\theta_n)\sin(t\lambda_n)\right]|n\rangle\otimes |e\rangle\cr
& &+ie^{-it(\mu-\omega_1/2)}\sin(2\theta_n)\sin(t\lambda_n)|n+1\rangle\otimes |g\rangle\cr
%%%%%
S_1|n+1\rangle\otimes |g\rangle&=&e^{itH_0}e^{-itH}|n+1\rangle\otimes |g\rangle\cr
&=&e^{itH_0}\left[\cos(\theta_n)e^{-itE_n^+}|n,+\rangle+\sin(\theta_n)e^{-itE_n^-}|n,-\rangle\right]\cr
&=&ie^{it(\mu-\omega_1/2)}\sin(2\theta_n)\sin(t\lambda_n)|n\rangle\otimes |e\rangle\cr
& &+e^{-it(\mu-\omega_1/2)}
\left[\cos(t\lambda_n)-i\cos(2\theta_n)\sin(t\lambda_n)\right]\cr
& &\times|n+1\rangle\otimes |g\rangle.\cr
& &
\ee
These expressions can be written as (\ref {S:step1}).

\section{Change in the state of the cold oscillator}

Here we calculate (\ref {transfer:en1}).

Note that $[a^\dagger a,G_1]=0$ and
$[a^\dagger a,a^\dagger Aa]=0$ and $[a^\dagger a,a^\dagger B]=a^\dagger B$
and $[a^\dagger a,a^\dagger Ca]=0$.
Using these relations one obtains
\be
[a^\dagger a,S]
&=&ie^{\frac i2\tau_1(\omega_1-2\mu)}
a^\dagger B\otimes E_+\otimes G_3\cr
& &+ie^{\frac i2\tau_3(\omega_3-2\delta)}e^{\frac i2\tau_1(\omega_1-2\mu)}
a^\dagger B\otimes E_+\otimes c^\dagger (Z-iV)c\cr
& &-e^{\frac i2\tau_3(\omega_3-2\delta)}e^{-\frac i2\tau_1(\omega_1-2\mu)}
Ba\otimes E_1\otimes c^\dagger Y\cr
& &+e^{-\frac i2\tau_3(\omega_3-2\delta)}e^{\frac i2\tau_1(\omega_1-2\mu)}
a^\dagger B\otimes E_2\otimes Yc\cr
& &-ie^{-\frac i2\tau_3(\omega_3-2\delta)}e^{-\frac i2\tau_1(\omega_1-2\mu)}
Ba\otimes E_-\otimes cc^\dagger (Z+iV).
\label {transfer:temp}
\ee
Note that
\be
X\equiv
cc^\dagger(Z^2+V^2)+Y^2=\sum_n\frac 1{n+1}|n\rangle\langle n|
\label {appC:id}
\ee
so that $cc^\dagger X=G_3+c^\dagger Xc=\Io$.
Hence, from (\ref {transfer:temp}) one obtains 
\be
D
&=&
S^\dagger [a^\dagger a,S]\cr
&=&
+\left[
ia^\dagger  aa^\dagger (A+iC)B\otimes E_+
+ aa^\dagger B^2\otimes E_2\right]
\otimes (G_3+c^\dagger Xc)\cr
& &
-\left[a^\dagger B^2a\otimes E_1
+i (A-iC) aa^\dagger Ba\otimes E_-\right]
\otimes cc^\dagger X\cr
&=&\bigg[aa^\dagger B^2\otimes E_2-a^\dagger B^2a\otimes E_1\cr
& &+ia^\dagger  aa^\dagger (A+iC)B\otimes E_+
-i(A-iC)B aa^\dagger a\otimes E_-\bigg]\otimes\Io.
\ee
This is (\ref {transfer:en1}).

% APPENDIX D
\section{Work performed during phases 1, 3, 4}

We first calculate $\Delta E_1=H_0-S_1H_0S_1^\dagger$. Note that one can write
$\Delta E_1=S_1[S_1^\dagger,H_0]$.
Therefore we start with calculating
\be
[S_1^\dagger,H_0]&=&-i(\omega_1-2\mu)e^{-\frac i2\tau_1(\omega_1-2\mu)}Ba\otimes E_-\cr
& &+i(\omega_1-2\mu)e^{\frac i2\tau_1(\omega_1-2\mu)}a^\dagger B\otimes E_+.
\ee
Now multiplying from the left with $S_1$ yields (\ref {work:phase1}).

%%%%%%%%%%%%%%%%%%%%%%%%%%%%%%%%%%%%%%%%%%%%%%%
Next calculate $\Delta E_3$ using
\be
\Delta E_3=S_3^\dagger H_0S_3-H_0
=S_3^\dagger [H_0,S_3].
\ee
One calculates
using  $[c^\dagger c,c^\dagger (Z- iV)c]=[c^\dagger c,cc^\dagger (Z+ iV)]
=[c^\dagger c,Y]=0$
\be
[H_0,S_3]
&=&\delta[E_3-E_2,S_3]+\omega_3[c^\dagger c,S_3]\cr
&=&i(\omega_3-2\delta) e^{\frac i2\tau_3(\omega_3-2\delta)}F_+c^\dagger Y
-i(\omega_3-2\delta) e^{-\frac i2\tau_3(\omega_3-2\delta)}F_-Yc.
\ee
 This gives using $F_-F_+=E_3,E_2F_+=F_+,F_+F_-=E_2,E_3F_-=F_-$
\be
S_3^\dagger [H_0,S_3]
&=&(\omega_3-2\delta)E_3Ycc^\dagger Y+i(\omega_3-2\delta)F_+c^\dagger (Z+iV)cc^\dagger Y\cr
& &-(\omega_3-2\delta)E_2c^\dagger Y^2 c -i(\omega_3-2\delta)F_-(Z-iV)cc^\dagger Yc.
\ee
It is then straightforward to obtain
\be
S_2^\dagger( S_3^\dagger H_0S_3-H_0)S_2
&=&(\omega_3-2\delta)E_2Ycc^\dagger Y-(\omega_3-2\delta)E_1c^\dagger Y^2 c\cr
& &-i(\omega_3-2\delta)E_+c^\dagger (Z+iV)cc^\dagger Y\cr
& &+i(\omega_3-2\delta)E_-(Z-iV)cc^\dagger Yc.\cr
& &
\ee
This yields $\Delta E_3$.

%%%%%%%%%%%%%%%%%%%%%%%%%%%%%%%%%%%%%%%%%%%%%%%
Finally calculate $\Delta E_4$.
One has using the simplified expression (\ref {cycles:step2simpl})
\be
S_2 H_0S_2^\dagger-H_0=2 [\mu+\delta,-\mu,-\delta].
\ee
Note that (using $E_2S_2=iE_-$, $F_+S_2=-iE_2$, and $E_3S_2=-iF_-$)
\be
S_3S_2
&=&\Io\otimes E_1S_2\otimes\Io+i\Io\otimes E_-\otimes G_3\cr
& &+e^{\frac i2\tau_3(\omega_3-2\delta)}\Io\otimes \left[
iE_-\otimes c^\dagger (Z-iV)c+E_2\otimes c^\dagger Y\right]\cr
& &+e^{-\frac i2\tau_3(\omega_3-2\delta)}\Io\otimes \left[
i F_-S_2\otimes Yc -iF_-\otimes cc^\dagger (Z+iV)\right].
\ee
This gives (using $S_2^\dagger E_1S_2=E_3$ and $S_2^\dagger E_2S_2=E_1$)
\be
S_2^\dagger S_3^\dagger E_1S_3S_2&=&E_3\cr
S_2^\dagger S_3^\dagger E_2S_3S_2&=&\Io\otimes E_1\otimes G_3+\Io\otimes E_1\otimes c^\dagger (Z^2+V^2)cc^\dagger c
+\Io\otimes E_2\otimes Y^2cc^\dagger\cr
& &-i\Io\otimes E_+\otimes c^\dagger(Z+iV)cc^\dagger Y\cr
& &+i\Io\otimes E_-\otimes Ycc^\dagger(Z-iV)c\cr
S_2^\dagger S_3^\dagger E_3S_3S_2&=&
\Io\otimes E_1\otimes c^\dagger Y^2 c+\Io\otimes E_2\otimes (Z^2+V^2)(cc^\dagger)^2\cr
& &+i\Io\otimes E_+\otimes c^\dagger(Z+iV)cc^\dagger Y\cr
& &-i\Io\otimes E_-\otimes Ycc^\dagger(Z-iV)c.
\ee
The result is
\be
\Delta E_4
&=&2(\mu+\delta)E_3\cr
& &-2\mu\Io\otimes \left[E_1\otimes G_3+E_1\otimes c^\dagger (Z^2+V^2)cc^\dagger c
+ E_2\otimes Y^2cc^\dagger\right]\cr
& &-2\delta\Io\otimes \left[E_1\otimes c^\dagger Y^2 c+E_2\otimes (Z^2+V^2)(cc^\dagger)^2\right]\cr
& &+2i(\mu-\delta)\Io\otimes \bigg[E_+\otimes c^\dagger(Z+iV)cc^\dagger Y
-E_-\otimes Ycc^\dagger(Z-iV)c\bigg].\cr
& &
\ee
Using $G_3+c^\dagger Xc=\Io$ and $cc^\dagger X=\Io$ this can be written as
\be
\Delta E_4
&=&-\Delta_2+2(\mu-\delta)\Io\otimes E_1\otimes c^\dagger Y^2 c-2(\mu-\delta)\Io\otimes E_2\otimes y^2cc^\dagger\cr
& &+2i(\mu-\delta)\Io\otimes \bigg[E_+\otimes c^\dagger(Z+iV)cc^\dagger Y
-E_-\otimes Ycc^\dagger(Z-iV)c\bigg].\cr
& &
\ee
This is (\ref {work:phase4}).

\section*{}
%%%%%%%%%%%%%%%%%%%%%%%%%%%%%

\end{document}